\documentclass[%
 reprint,
 amsmath,amssymb,
 aps,
]{revtex4-1}

\usepackage{graphicx}
\usepackage{dcolumn}
\usepackage{bm}

\begin{document}

\preprint{APS/123-QED}

\title{Fibre Inflation and Reheating}

\author{Paolo Cabella}
\email{paolo.cabella@roma2.infn.it}
\author{Alessandro Di Marco}
\author{Gianfranco Pradisi}
\affiliation{%
University of Rome - Tor Vergata and INFN - Sezione di Roma ``Tor Vergata"\\
Via della Ricerca Scientifica 1, 00133 Roma, Italy}%


\begin{abstract}
We present constraints on the reheating era within the string Fiber Inflation scenario,
in terms of the effective equation-of-state parameter of the reheating fluid, $w_{reh}$.
The results of the analysis, completely independent on the details of the inflaton physics around the vacuum, illustrate the behavior of the number of $e$-foldings during the reheating stage, $N_{reh}$, and of the final reheating temperature, $T_{reh}$,  as functions of the scalar spectral index, $n_s$.
We analyze our results with respect to the current bounds given by the PLANCK mission data and to upcoming cosmological experiments.
We find that large values of the equation-of-state parameter ($w_{reh}>1/3$) are particularly favored as the scalar spectral index is of the order of $n_s\sim 0.9680$, with a $\sigma_{n_s}\sim 0.002$ error.
Moreover, we compare the behavior of the general reheating functions $N_{reh}$ and $T_{reh}$ in the fiber Inflation scenario with the one extracted by the class of the $\alpha$-attractor models with $\alpha=2$.  We find that the corresponding reheating curves are very similar in the two cases.

\end{abstract}

\pacs{Valid PACS appear here}
\maketitle


\section{Introduction}

The single field slow-roll inflation is among the most promising models for a description of the early universe.
This scenario involves, in its simplest version, an homogeneous, minimally coupled to gravity and canonically normalized scalar field $\varphi$, called inflaton \cite{1}.
The scalar field is subjected to a potential $V(\varphi)$ that in the slow-roll phase can drive an accelerated expansion of the Universe,
with a typical duration of $N_*\sim 60$ $e$-folds.
Nevertheless, several features of the inflationary physics are still elusive, and one of these is represented by the reheating phase in which
the energy stored in the inflaton field is converted to thermal radiation, giving rise to the early radiation-dominated epoch of the standard hot big bang model
\cite{2,3,4,5,6,7,8,9,10,11,12,13,14,15}.
The reheating is a fundamental stage of the Universe evolution and its details affect and constrain the range of several observables, for instance $n_s$ and $r$.
Moreover, it is strongly model dependent and can be highly non-trivial, due to the nonlinear and nonperturbative processes involved.
However, it is always possible and convenient to characterize this stage by two main parameters: the duration of the epoch, $N_{reh}$,  and the final temperature, $T_{reh}$.
Both these quantities depend on the equation-of-state parameter $w_{reh}$ (EoS), in an effective description based on a cosmic fluid.
In this paper, we want to investigate the constraints on $N_{reh}$ and $T_{reh}$ as functions of the scalar spectral index, $n_s$,
focusing on the class of the so-called string fiber inflation models \cite{16,17,18}.  It results in a large volume scenario \cite{LVS} where the inflaton is a K\"ahler module, and it typically predicts a scalar spectral index $n_s\sim 0.97$, together with a
tensor-to-scalar ratio of the order of $r\sim 0.006$ for an inflationary number of $e$-foldings $N_*\sim 60$.
In our analysis, we highlight the kind of fiber-inflation reheating phases that are currently favored by the PLANCK mission results \cite{20} and also the emerging scenarios that appear to be more interesting 
in view of the next generation of cosmological experiments \cite{21,22,23,24,25,26}, in terms of the mean value of the parameter $w_{reh}$.
In addition, we also compare the post-inflationary phase of the fiber inflation scenario with the one coming from the 
$\alpha$-attractor class of models (for details, see references \cite{27,28,29,30,31,32}) with $\alpha=2$, which owns a potential with an almost identical shape during inflation.  
The paper is organized as follows.  In Section II, we give a sketch of the main properties of the fiber inflation class of models.
In Section III, we briefly review the description of the reheating phase, stressing the relation between $N_*$ and $N_{reh}$ in terms of the $w_{reh}$ parameter.  We also give details on how we perform our analysis.  In Section IV we report and describe the numerical results.  We also discuss their impact on the fiber inflation scenario and the comparison with the $\alpha$-attractor (with $\alpha=2$) model.
In this paper (unless explicitly indicated) we use units in which the reduced Planck mass is set to $M_p=1$.

\section{Fibre Inflation Models}

\begin{figure}[htbp]
\centering
\includegraphics[width=8.5cm, height=6cm]{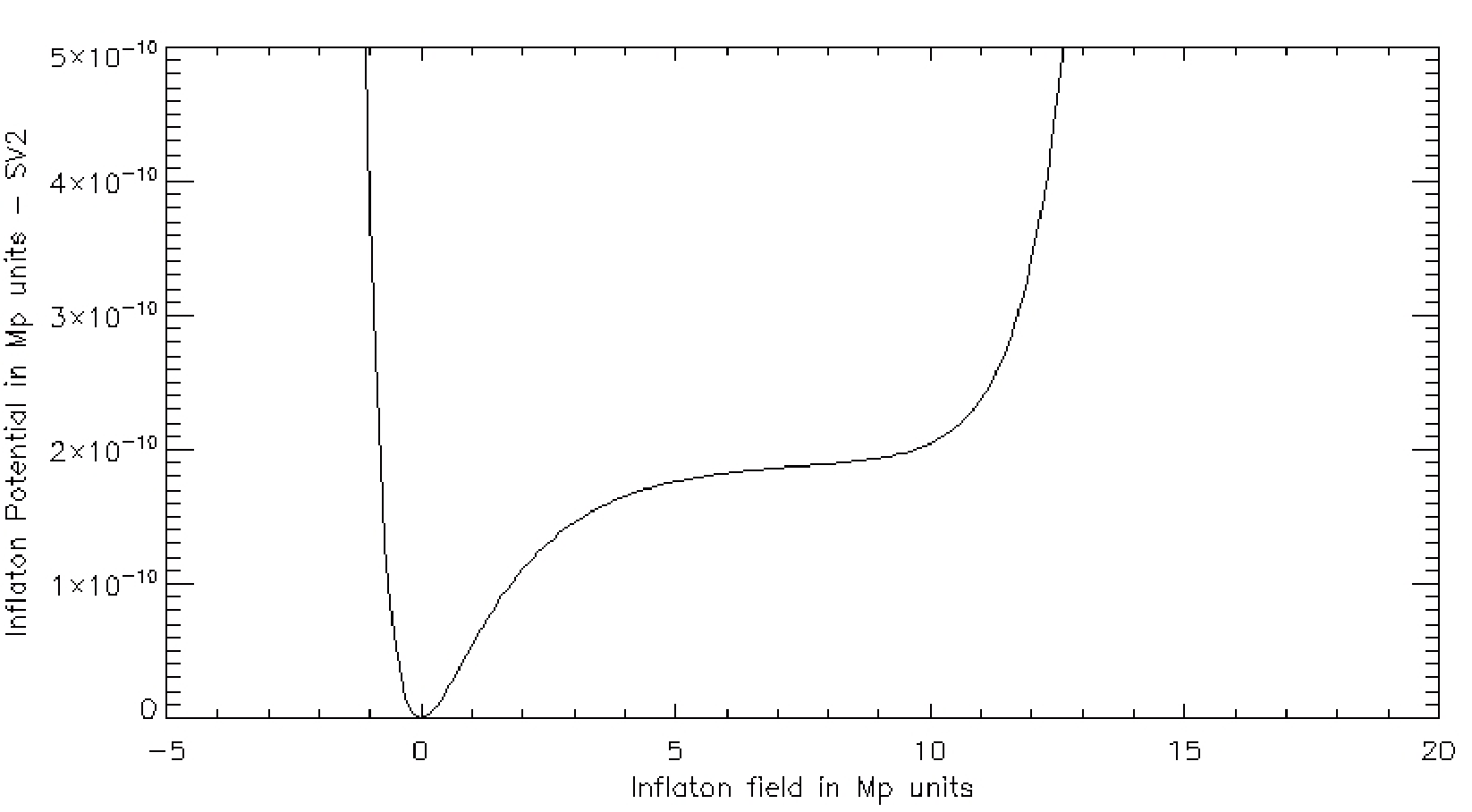}
\caption{Inflationary potential in the fiber inflation scenario in $M_p$ units. As one can see, the potential grows exponentially for value of the scalar field
$\varphi>10$ in $M_p$ units}
\label{fig: fibre_potential_sv1}
\end{figure}

The inflationary process may have occurred at very high-energy scales and
this suggests that the scalar field could be an effective degree of freedom in the low-energy limit of some UV consistent model of quantum gravity.  The best candidate to describe physics in this regime is string theory. 
Thus, it appears natural to analyze realizations of inflation within string theory \cite{BauMcA}. In this context, the main drawback related to string-derived or string-inspired models of inflation is notoriously connected to the presence of potentially large quantum corrections that could destroy the naive flatness of the semiclassical inflaton potential, obtained using the low-energy effective supergravity approximation. As mentioned, we focus on a class of promising string models called fiber inflation \cite{16,17,17}, where quantum corrections are under quite robust control and predictions are in agreement with the experimental data, as we now  briefly review.  
The starting point of fiber inflation is to consider an orientifold \cite{AngSag} of Type IIB compactified on a six-dimensional Calabi-Yau (CY) manifold \cite{CY} resulting as an elliptic ($T^4$ or $K3$) fibration over a $\mathbb{P}^1$ base, with a globally consistent ({\it i.e.} without tadpoles) configuration of D3/D7 branes.  Several explicit examples are known of this kind of realizations \cite{18}. To stabilize the complex structure moduli and the dilaton, one has to assume that background fluxes are also turned on \cite{RevFlux}. To get a viable model, additional requirements are needed. First of all, there must be a hierarchy of masses in such a way that only one module remains as a good candidate for the inflaton. This is obtained as follows: the $h^{1,1}$ \cite{37} 
 K\"ahler moduli of which the real part is basically the volume of the corresponding cycle inside the CY, are of two kinds: the blowup ones, local and related to exceptional divisors realizing a ``resolution'' of the singular points, and the remaining ``big'' K\"ahler moduli, which include also the proper volume of the fiber. In the large-volume scenario, namely in the regime where the global volume of the internal CY is exponentially large, the semiclassical (tree-) level potential for the K\"ahler moduli is exactly vanishing, due to the so-called no-scale structure of the K\"ahler potential.  The inclusion of quantum corrections is crucial in order to get all moduli stabilized and a potential driving inflation for the candidate inflaton \cite{CCQ}. Three types of corrections must be considered: those related to the ``massive'' string states (higher derivative $\alpha'$ corrections), the ones related to the string loops ($g_s$ corrections, where $g_s$ is the string coupling constant of which the value depends on the vacuum expectation value of the stabilized dilaton) and, finally, nonperturbative corrections to the superpotential, related for instance to euclidean D-brane configurations and/or gaugino condensation. Some of these quantum effects have been explicitly calculated within string theory \cite{loops}, some others have been guessed by comparison with low-energy known effects, like the Coleman-Weinberg mechanism \cite{jump,CCQ}. Quantum corrections generate a potential for the moduli.  In particular, the CY volume and the small K\"ahler moduli are stabilized at large values. The small next-to-leading corrections, on the other hand, give rise to the potential for the semiclassically flat moduli, providing at the same time a hierarchy of masses with respect to the volume and the small moduli. To get single-field inflationary models, one has to restrict to compactifications with only one semiclassically flat direction, module of which is the natural candidate for the inflaton. The resulting potential can be described as follows. In the simplest case, the inflaton is related to the fiber volume, namely to the exponential of a canonically normalized scalar field. To be precise, the inflaton $\varphi$ is identified with the displacement of the scalar field from the minimum of the potential that can be written, with a judicious analysis of the coefficient \cite{41} in the following form:
\begin{equation}\label{potent}
V(\varphi)=V_0 \left( c_0 + c_1 e^{- k \varphi/2}  + c_2 e^{- 2 k \varphi}  + c_3 e ^{k \varphi} \right) ,
\end{equation}
where
\begin{align}
c_0 &=3-M ,\nonumber \\
c_1&= -4\left( 1+\frac{M}{6}  \right) , \nonumber\\
c_2&= \left( 1+\frac{2M}{3}  \right) , \nonumber\\
c_3&= M .\nonumber
\end{align}
In Eq.\eqref{potent} $k=2/\sqrt{3}$ and $V_0$ depend upon the CY volume.
In Fig.($\ref{fig: fibre_potential_sv1}$) we plot the potential function.
It should be noticed that $M\sim g_s^4 <<1$, so the potential can drive inflation. In the plateau, corresponding to large values of $\varphi$, it can be approximated by
\begin{equation}
V(\varphi)\simeq V_0 \ \left[ 3 - 4 e^{-  \varphi/\sqrt{3}}   \right]  .
\end{equation}
The overall (adimensional) normalization
can be written in the form \cite{16}
\begin{eqnarray}\label{eqn: overall}
V_0 = \frac{\mathcal{C}_2}{<\mathcal{V}>^{10/3}}  ,
\end{eqnarray}
where $\mathcal{C}_2$ is related to the quantum corrections, thus depending on the stabilized moduli, while $\mathcal{V}$ is the CY volume measured in string scale units.  Clearly, in order to compare with experimental data, one has to use physical units for $V_0$ \cite{16,42}.

\section{Reheating and equation of state}

\begin{figure}[htbp]
\centering
\includegraphics[width=8.5cm, height=6cm]{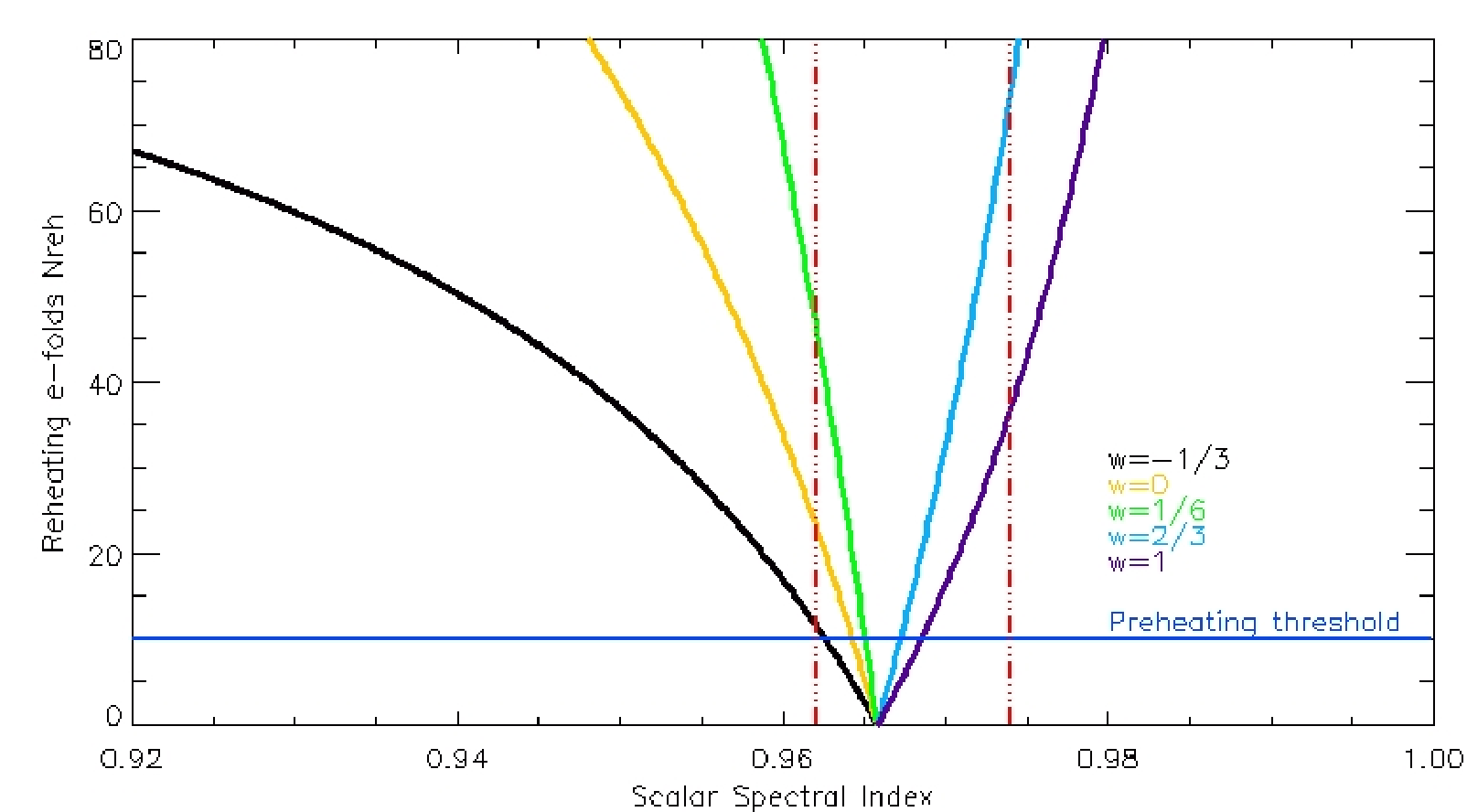}
\caption{Beheaviour of the $N_{reh}(n_s)$ function in the case of fibre inflation for different value of EoS. The red vertical dashed and dotted lines represent
the current $1$-$\sigma$ value on the scalar spectral index given by the PLANCK mission, $\sigma_{n_s}=0.006$}
\label{fig: reh_fibre}
\end{figure}

\begin{figure}[htbp]
\centering
\includegraphics[width=8.5cm, height=6cm]{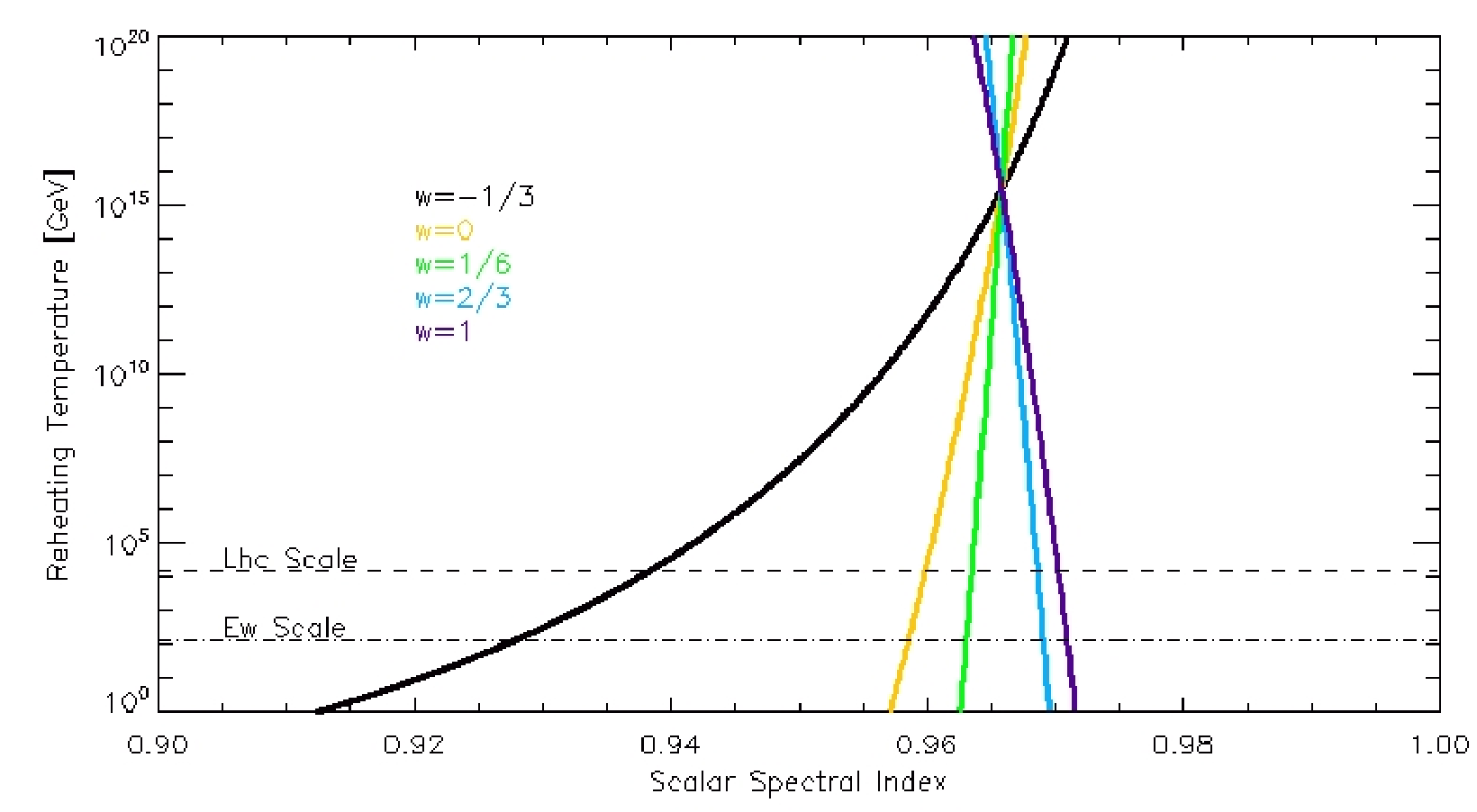}
\caption{Beheaviour of the $T_{reh}$ function in the case of fibre inflation for different values of EoS. The black horizontal dashed lines represent two 
fundamental physical scales: the energy scale
at which LHC currently works and the electroweak scale.}
\label{fig: temp_fibre}
\end{figure}

\begin{figure}[htbp]
\centering
\includegraphics[width=8.5cm, height=6cm]{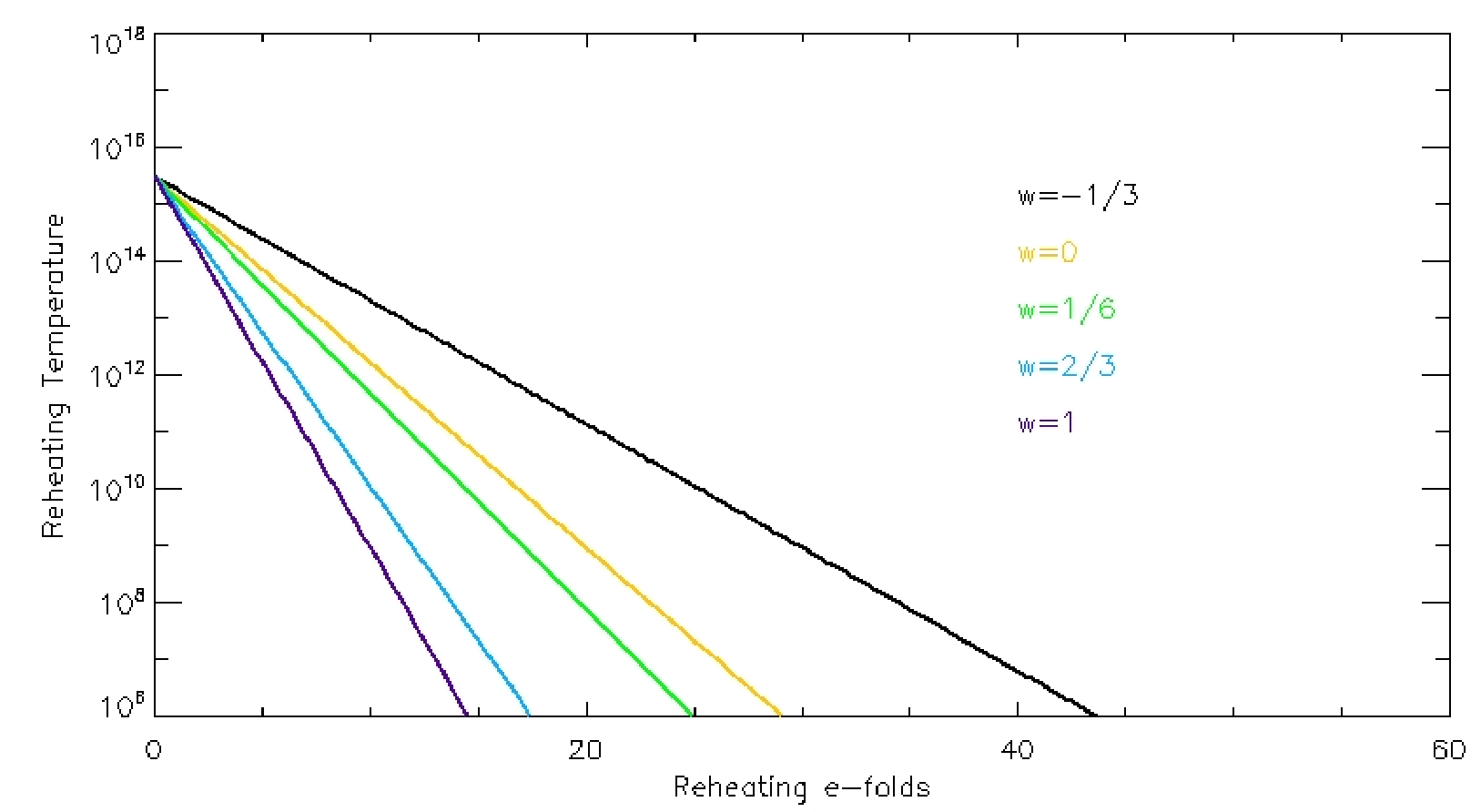}
\caption{Reheating temperature (log scale) as function of the number of $e$-foldings of the reheating epoch for five different values of the equation of state parameter.}
\label{fig: NvsT}
\end{figure}

\begin{figure}[htbp]
\centering
\includegraphics[width=8.5cm, height=6cm]{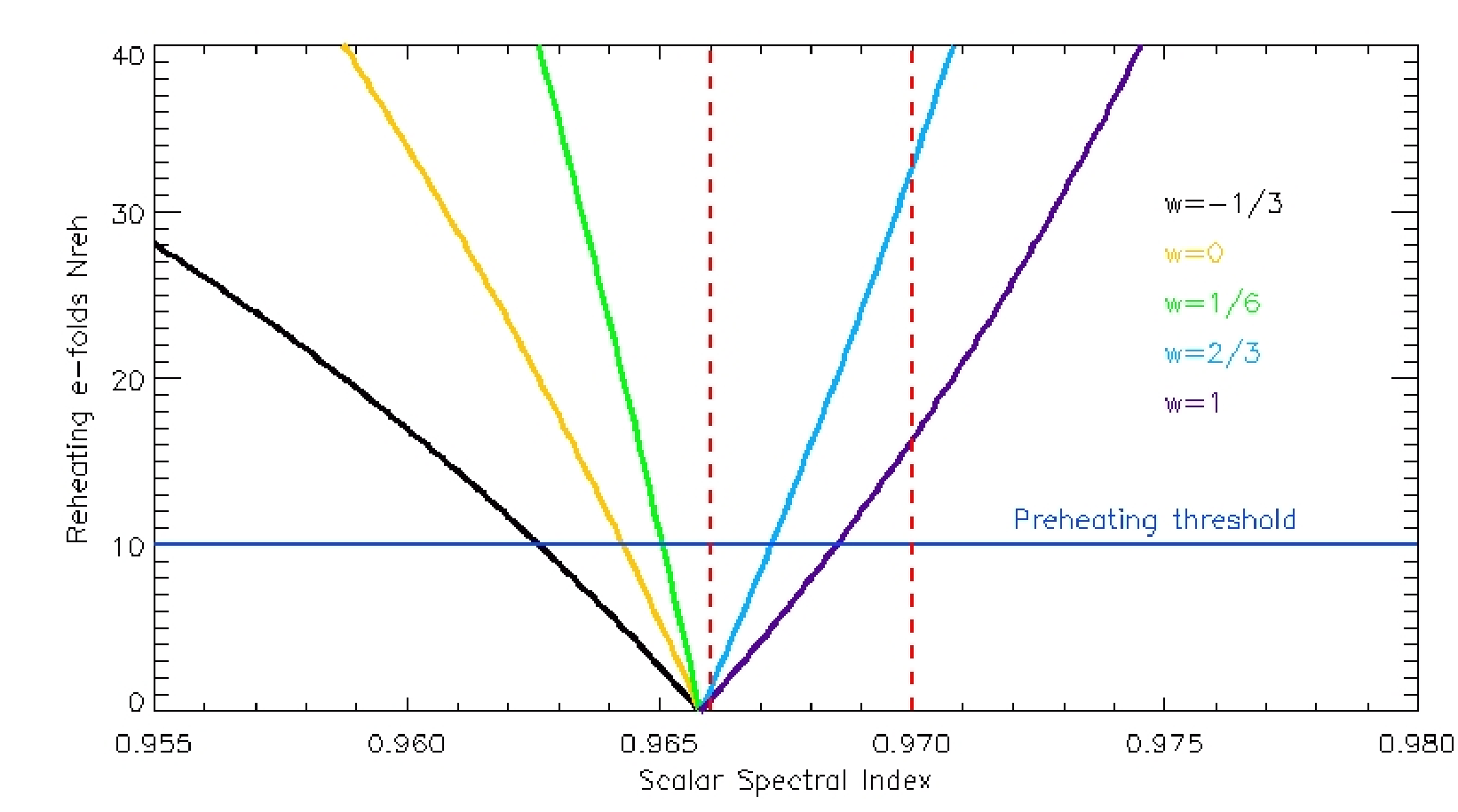}
\caption{Sketch on the $N_{reh}(n_s)$ function around the mean value of a future detection with a mean value $n_s=0.9680$ with $1$-$\sigma_{n_s}\sim 0.002$.
In this case, low values of the EoS ($w_{reh}<1/3$) would be disfavored. On the contrary, larger values ($w_{reh}>1/3$) would be favored and allow pre-heating phases.}
\label{fig: zoom1}
\end{figure}

\begin{figure}[htbp]
\centering
\includegraphics[width=8.5cm, height=6cm]{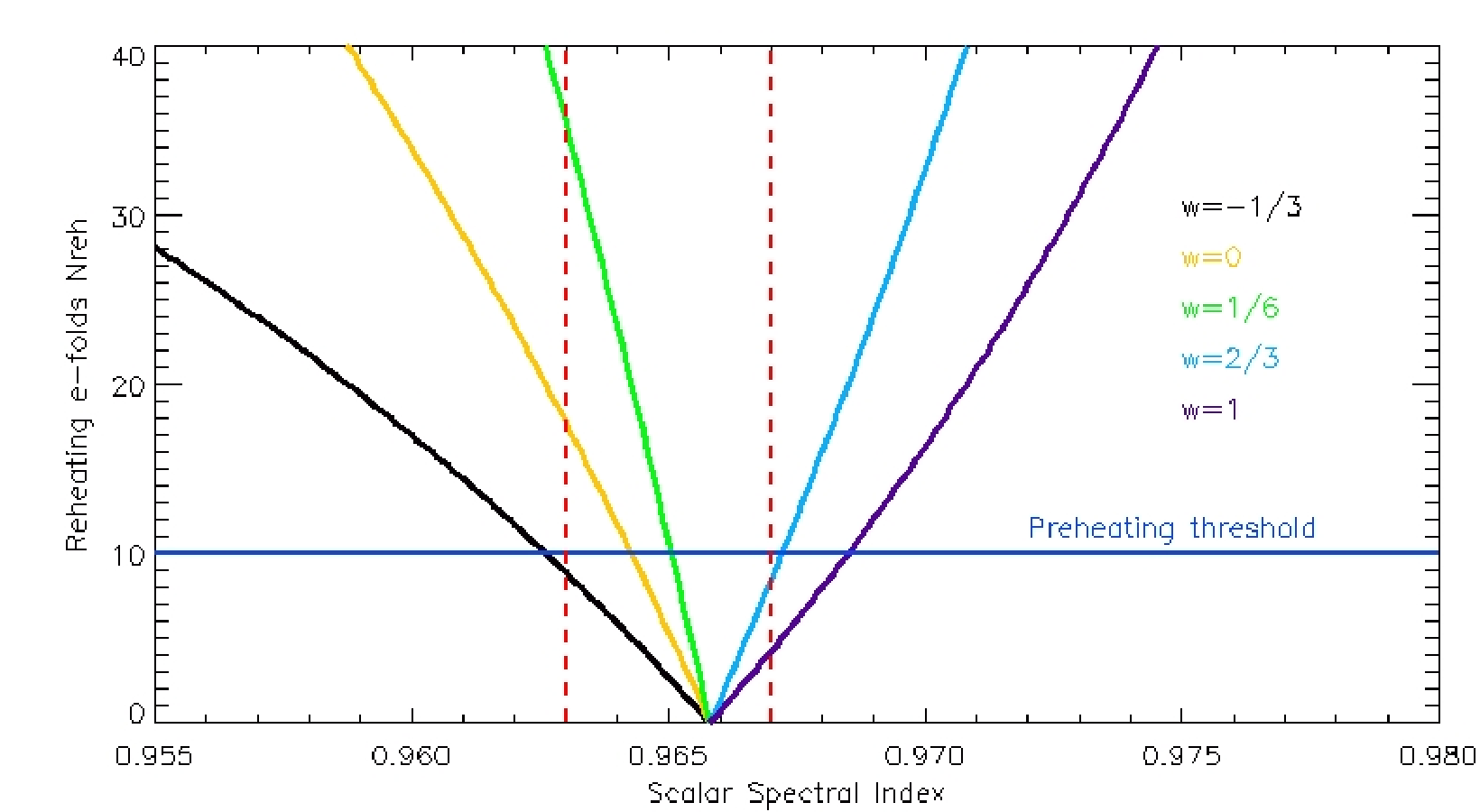}
\caption{ketch on the $N_{reh}(n_s)$ function around the mean value of a future detection $n_s=0.9650$ with $1$-$\sigma_{n_s}\sim 0.002$. In this case, low values 
of the EoS ($w_{reh}<1/3$) would be favored and can allow prolonged reheating phase. Meanwhile, larger values of the EoS ($w_{reh}>1/3$) tend to be not favored.}
\label{fig: zoom2}
\end{figure}

Inflation comes to end when the inflaton field reaches the value $\varphi_{end}$ and the potential start to steepen. After that, the inflaton field falls in the minimum of the potential $V(\varphi)$, acquires a mass $m_{\varphi}$, oscillates, and decays producing the relativistic degrees of freedom of radiation-dominated plasma.
This reheating phase can be modeled using an effective cosmic fluid described in terms of an equation-of-state parameter $w_{reh}$ \cite{13,14,15}.  The first model of reheating was proposed in \cite{2}, in a relatively simple setting where the field coherently oscillates around the minimum of a quadratic potential $\sim m^2\varphi^2$.
Consequently, a cold gas of $\varphi$-bosons that
decay into relativistic particles is produced. Because of mutual interactions, the particles reach a thermal equilibrium and give rise to a graceful exit toward the radiation-dominated epoch.  More details can be found in references \cite{3,4,5,6,7}.  However, it is a common belief that the physics of reheating can involve far more complicated mechanisms, as suggested by several authors \cite{8,9,10,11,12}.  
For instance, it is customary to have different decay rates for different particles and to expect that nonequilibrium phenomena, nonlinearities and turbulence could play a role.  Fortunately, in a first approximation one can always define two quantities that characterize the reheating mechanisms in a completely general way.
The first one is the number of $e$-foldings during reheating, $N_{reh}$, which parametrizes the duration of the reheating stage itself. The second is the reheating temperature, $T_{reh}$, which represents the energy scale of the Universe at which the reheating is completely realized.  In particular, the inflationary number of $e$-foldings depends on the complete cosmic history of the Universe as 
\begin{eqnarray}\label{eqn: efolds}
N_*=-\ln\left(\frac{k_*}{a_0 H_0} \right) - N_{reh} - N_{pr} + \ln\left( \frac{H_*}{H_0}\right)  ,
\end{eqnarray} 
where $k_*$ is the pivot scale stretched by the expansion at early times; $a_0$ and $H_0$ are the scale factor and the Hubble rate at current epoch, respectively;
$H_*$ is the Hubble rate during inflation; $N_{reh}$ is the number of $e$-foldings during the reheating stage and $N_{pr}$ the number of $e$-foldings
during the subsequent postreheating phases.
Eq.($\ref{eqn: efolds}$) shows how the number $N_*$ as well as other observables like $n_s$ and $r$ are strongly dependent on $N_{reh}$.
Inverting the relation in Eq.($\ref{eqn: efolds}$) with respect to $N_{reh}$ one gets 
\begin{eqnarray}\label{eqn: reh}
N_{reh}=\frac{4}{1-3 w_{reh}}f(\beta_i,O_i,N_*) ,
\end{eqnarray}
where the function $f$ comes out to be 
\begin{align}
f(\beta_i,O_i,N_*)&=\left[ -N_* - \ln\left( \frac{k_*}{a_0 H_0} \right) + \ln\left( \frac{T_0}{H_0} \right) \right] \nonumber\\
&+\left[ \frac{1}{4}\ln\left( \frac{V^2_*}{M^4_p \rho_{end}} \right) -\frac{1}{12}\ln(g_{reh})  \right] \nonumber\\
&+\left[ \frac{1}{4}\ln\left(\frac{1}{9}\right) + \ln\left( \frac{43}{11}\right)^{\frac{1}{3}} \left(\frac{\pi^2}{30} \right)^{\frac{1}{4}}\right] .
\end{align}
The quantities $\beta_i$ are the parameters of the considered model of inflation while $O_i$ represents the known cosmological parameters.
In particular, $T_0$ is the cosmic macrowave background (CMB) photon temperature, $V_*$ is the vacuum energy density at  
the horizon exit, $\rho_{end}$ is the energy density when inflation stops and $g_{reh}$ indicates the number of degrees of freedom of relativistic species when reheating comes to the end.
In addition, one can show that the reheating temperature is defined as
\begin{eqnarray}\label{eqn: gen_temp}
T_{reh}=\left( \frac{40 V_{end}}{\pi^2 g_{reh}} \right)^{1/4} \exp{ \left[-\frac{3}{4}(1+w_{reh})N_{reh}\right]} .
\end{eqnarray}
Thus, information on $N_*$ (and consequently on cosmological observables) can be translated into information on $N_{reh}$ and $T_{reh}$.
We want to exploit these relations to extract constraints about the reheating phase in the fiber inflation scenario, corresponding to different values of 
the equation-of-state parameter $w_{reh}$.
This procedure was already applied to power-law potentials by Dai et al. in Ref. \cite{13} (for similar andd further constraints, see also Ref. \cite{14}) and to $\alpha$-attractor models by Ueno and Yamamoto in Ref. \cite{15}.  Our analysis is also devoted to extract further possible constraints on the reheating phase in fiber inflation models given by the next generation of CMB polarization and Gravitational Waves (GW) experiments \cite{21,22,23,24,25,26}.
Let us start by properly rewriting Eq.($\ref{eqn: reh}$). For this, it is convenient to set $k_*=0.002$ Mpc$^{-1}$, $H_0=1.75\times 10^{-42}$ GeV, $T_0= 2.3\times 10^{-13}$ GeV and $g_{reh}=100$, in such a way that
\begin{eqnarray}
N_{reh}=\frac{4}{1-3 w_{reh}}\left[\xi_0-N_* + \frac{1}{4}\ln\left( \frac{V^2_*}{M^4_p \rho_{end}}\right)\right],
\end{eqnarray}
where $\xi_0=64.24$. It should be underlined that the inflationary number of $e$-foldings at first order in the observables results to be
\begin{eqnarray}
N_*\sim\frac{2}{1-n_s}
\end{eqnarray}
within the fiber inflation scenario. 
Moreover, the energy piece is written using the full form of the fiber inflation potential exploiting the SV2 set of parameters (see the original paper \cite{16})
with $\mathcal{C}_2 \sim 10^4$, $\mathcal{V} \sim 1.6 \times 10^3$,
$M \sim 2.3 \times 10^{-6}$ and with an energy scale of inflation of order $6.8 \times 10^{15}$ GeV.
The quantities $N_{reh}$ and $T_{reh}$ as functions of the scalar spectral index $n_s$ are reported in Fig.($\ref{fig: reh_fibre}$) and Fig.($\ref{fig: temp_fibre}$),
respectively.
We plot $N_{reh}$ and $T_{reh}$ in terms of the five different values $w_{reh}=-1/3,0,1/6,2/3,1$ of the equation-of-state parameter.  It should be noticed that the range $w_{reh}<1/3$ is favored by Quantum Field Theory (QFT), while $w_{reh}>1/3$ is quite unnatural from the QFT point of view, because it requires a potential that behaves like $\sim\varphi^n$, with $n>6$, around the minimum.
In Fig.($\ref{fig: NvsT}$) we show the behavior of $T_{reh}$ as a function of $N_{reh}$.  This relation is useful to read the energy scale at the end of reheating with respect to its time duration, for each considered $w_{reh}$.
In Fig.($\ref{fig: zoom1}$) we show the favored reheating phase with respect to a future detection of the scalar spectral index
with $n_s=0.9680$ (consistently with PLANCK current results) and $\sigma_{n_s}=0.002$.
On the other hand, in Fig.($\ref{fig: zoom2}$) we report the same investigation, assuming that a future experiment will displace the mean value of the scalar index
to $n_s=0.9650$ with the same $\sigma_{n_s}$.


\section{Numerical results and discussion}

\begin{figure}[htbp]
\centering
\includegraphics[width=8.5cm, height=6cm]{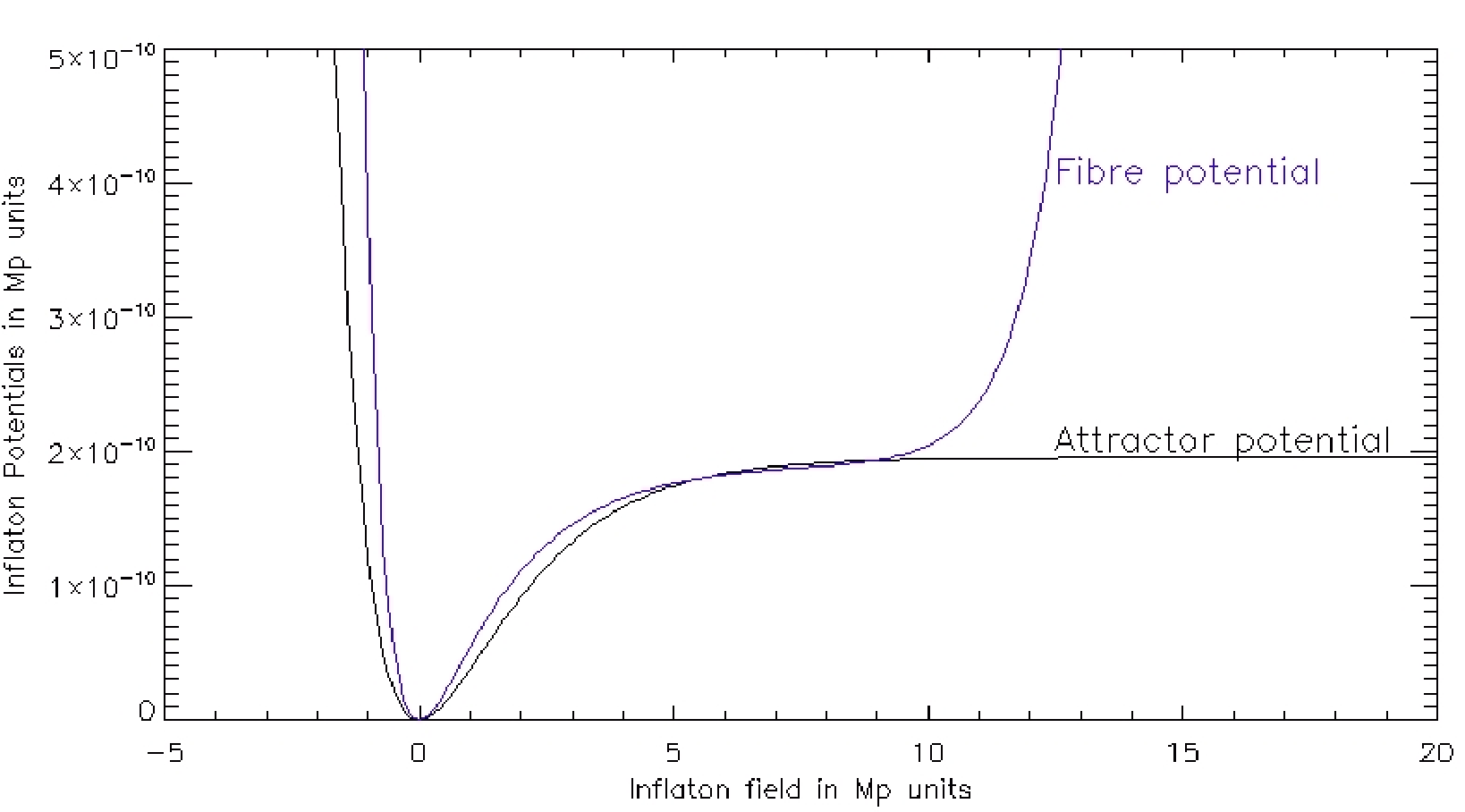}
\caption{Fibre Inflation potential and $\alpha$-attractor potential with $\alpha=2$. The shape of the two functions in the inflationary phase are very similar.
 The value of the inflaton field at horizon crossing moment is $\varphi_*\sim 5.7$ where as the value at the end of inflation is $\varphi_{end}\sim 1$ for $N_*\sim 60$.}
\label{fig: fibreVSalpha}
\end{figure}

\begin{figure}[htbp]
\centering
\includegraphics[width=8.5cm, height=6cm]{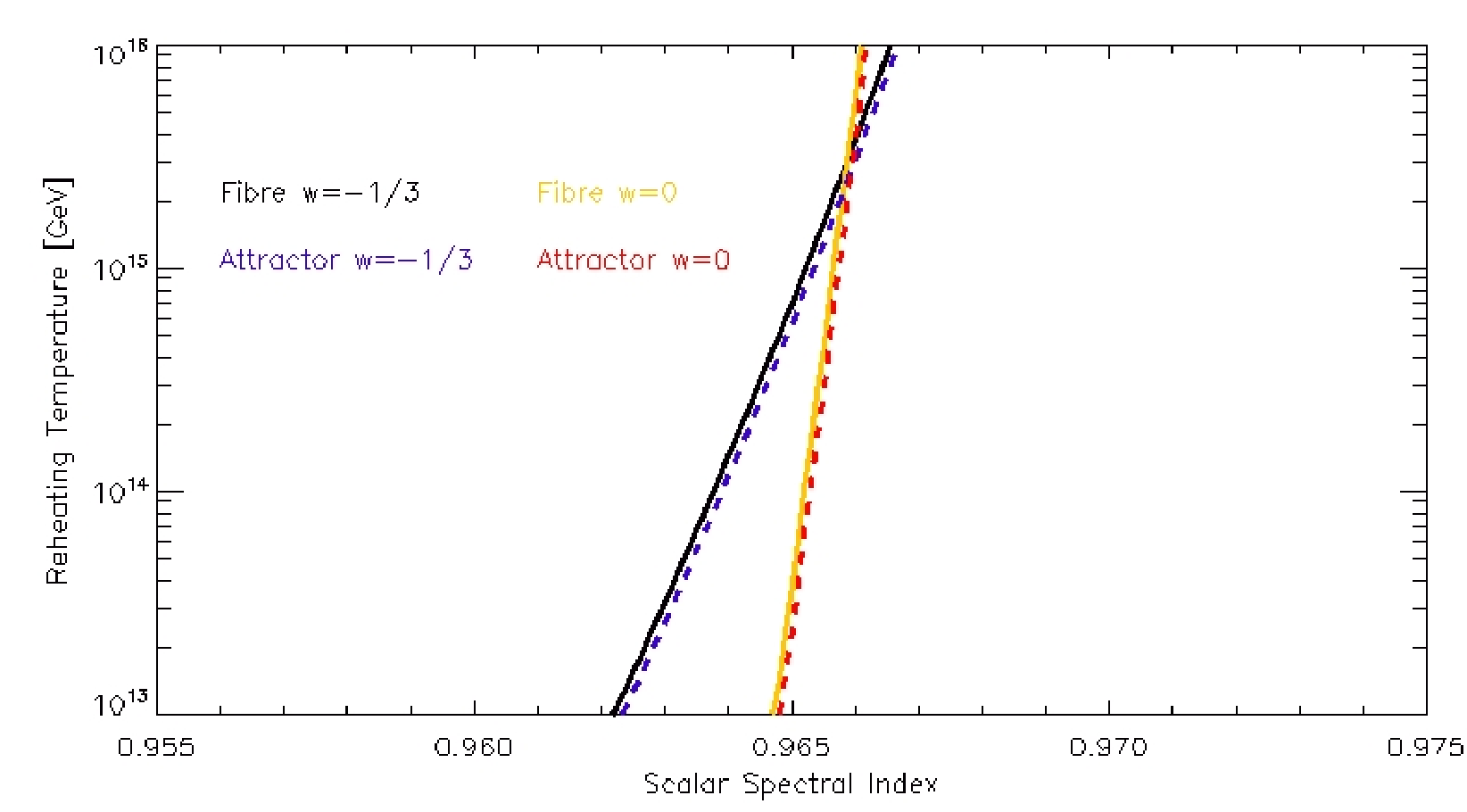}
\caption{Sketch on the reheating temperature as function of $n_s$ for Fibre Inflation and $\alpha=2$-attractor model.
The similarity of the two functions reveals that the postinflationary reheating phase are quit similar. The degeneracy could be broken by analyzing the 
preheating properties of the two models.}
\label{fig: fibreVSalpha_temp}
\end{figure}

The analysis performed in the previous section shows important facts.  
The current cosmological bounds on the scalar spectral index $n_s$ suggest that by and large
all the considered fiber reheating phases are allowed within PLANCK data constraints.
In particular, Fig.($\ref{fig: reh_fibre}$) shows that the lowest value for the EoS ($w_{reh}=-1/3$) is consistent with the current bounds if $N_{reh}<15$ with a corresponding reheating temperature $T_{reh}<10^{13}$ GeV.
This implies a short reheating epoch with a not excluded preheating stage.
The simplest scenario for the EoS, $w_{reh}=0$, which implies a matter dominated expansion of the reheating Universe, requires $N_{reh}<25$ to 
be in agreement with the current PLANCK results and $T_{reh}<10^8$ GeV.  On the other hand, equations-of-state parameter $w_{reh}>1/6$ induces a lower reheating temperature.
It is obvious that a more precise measurement of $n_s$ is crucial in order to discriminate between the possible scenarios. 
Let us consider, for instance, a possible future detection in which $n_s=0.9680$ with $\sigma_{n_s}=0.002$, as shown in Fig.($\ref{fig: zoom1}$).  
In this case, $w_{reh}=-1/3,0,1/6$ would be strongly disfavored.  
On the other hand, a value $w_{reh}=2/3$ would be particularly favored.  
On the contrary, if future experiments will provide a lower value of the scalar spectral index like $n_s=0.9650$ (Fig.($\ref{fig: zoom2}$)) keeping the same uncertainty, larger values of the equation-of-state parameter (like $w_{reh}>1/3$) are disfavored. 
In particular, $w_{reh}=-1/3$ turns out to be in agreement with the absence of a preheating phase, while scenarios with $w_{reh}=0,1/6$ are both consistent with the observations for large values of $N_{reh}$ as well, allowing a prolonged preheating stage. 
It should be noticed that these results strongly depend on the overall normalization of Eq.($\ref{eqn: overall}$).
It is worth to compare our results with those related to the very interesting class of $\alpha$-attractor inflationary models (see Sec. I for references).  
The scalar potential of the exponential version of this class (the so-called E-model) is given by 
\begin{eqnarray}
V(\varphi)=\Lambda^4\left(1-e^{-b\varphi/M_p}\right)^2 ,
\end{eqnarray}
with
\begin{eqnarray}
b=\sqrt{\frac{2}{3\alpha}} , \quad \quad R_K=-\frac{2}{3\alpha} ,
\end{eqnarray}
with $R_K$ being the curvature of the inflaton scalar manifold.
It is widely known that $\alpha$ attractors reproduce the inflationary shape of several models of inflation like the Goncharov-Linde model ($\alpha=1/9$) \cite{43}
the Starobinsky model ($\alpha=1$) \cite{44,45} and the Higgs inflationary model \cite{46}.  It should be noticed that the fiber inflation model behaves in a very similar way to the $\alpha$-attractor model with $\alpha=2$, at least in the ``plateau'' regime.  This fact can be well appreciated by comparing the shapes of the potential for the two models, both plotted in Fig.(\ref{fig: fibreVSalpha}).  
The general properties of the reheating stage of $\alpha$ attractors have been very well studied by Ueno and Yamamoto in Ref. \cite{15}.  
In Fig.(\ref{fig: fibreVSalpha_temp}) we report quantitative results about the reheating temperature in both the fiber inflation and the corresponding $\alpha=2$-attractor model, for the values $w_{reh}=-1/3$ and $w_{reh}=0$ of the equation of state parameter.  
Despite the different shape of the potentials, the temperature curves are very similar in the two cases, suggesting a very close postinflationary cosmic history
with analogous predictions about the values of $n_s$ and $r$.
A natural question is how these two models can be discriminated. 
In fact, the shapes of the two scalar potentials during the inflationary expansion is quite similar, as we have already seen.\\
This leads to similiar values for the vacuum expectation value of the scalar field at the time of the horizon crossing
and at the end of inflation.
Thus, in order to
properly label the two models, we should focus on regions of the\\
exploited potentials that are
\begin{itemize}
\item cosmologically relevant
\item significally different
\end{itemize}
One way could be to choose regions around the minimum of the potentials where the functions do not coincide (except for very tiny values of $\varphi$). 
The form of the potential about the minimum is crucial to determine the production of an additional diffuse background of GW due to possible preheating effects. This background would be characterized by sharp peaks at very high frequencies. 
Symmetric models with $\varphi^2$ or $\varphi^4$ potentials were studied especially by Kofmann et al. in \cite{5}.
More recently, analysis about non symmetric models have been proposed in \cite{47}, where it is shown how the peaks in the GW spectrum of these non symmetic models should exceed the standard preheating spectrum. 
Scalar potentials like those of $\alpha$ attractors or fiber inflation models are not symmetric around the minimum in $\varphi=0$, also for $\varphi<M_p$.
The expansion around $\varphi=0$ of the $\alpha$ attractor can be written 
\begin{equation}
V(\varphi)\simeq \frac{1}{2}m_{(\alpha)}^2 \varphi^2 - \frac{1}{3!}g_{(\alpha)} \varphi^3 + \frac{1}{4!} \lambda_{(\alpha)} \varphi^4
\end{equation}
where
\begin{equation}
m_{(\alpha)}^2= 2b^2\Lambda^4, \quad g_{(\alpha)}=6b^3\Lambda^2, \quad \lambda_{(\alpha)}=14b^4\Lambda^4,
\end{equation}
while the corresponding result for the fibre inflation potential is
\begin{equation}
V(\varphi)\simeq \frac{1}{2}m_{(f)}^2 \varphi^2 - \frac{1}{3!}g_{(f)} \varphi^3 + \frac{1}{4!} \lambda_{(f)} \varphi^4,
\end{equation}
with
\begin{eqnarray}
m_{(f)}^2 &=& V_0 k^2 \left( \frac{c_1}{4} + 4c_2 + c_3 \right),\\
g_{(f)} &=& V_0 k^3 \left( \frac{c_1}{8} + 8c_2 - c_3 \right),\\
\lambda_{(f)} &=& V_0 k^4 \left( \frac{c_1}{16} + 16c_2 + c_3 \right) . 
\end{eqnarray}
In both cases, the third-order term measures the antisymmetry
of the potentials.
Because of
\begin{equation}
V_0=\Lambda^4 \quad \mbox{and}  \quad k=2b,
\end{equation}
the ratio between the third-order terms results in
\begin{equation}
\frac{g_{(f)}}{g_{(\alpha)}}=\frac{4}{3}\left( \frac{c_1}{8} + 8c_2 - c_3 \right)\simeq 10 .
\end{equation}
Therefore, one can expect that this level of discrepancy could induce appreciable differences in the structure of GW preheating peaks and consequently resolve the degeneracy.
One can also generalize fiber inflation models by including higher order $\alpha'$ corrections.
They have recently calculated \cite{48} and turn out to be comparable with the $g_s$ corrections, giving rise to a potential with a shape similar to the ones discussed
in this paper. It would be interesting to extend our analysis to these models as well.

\begin{acknowledgments}
This work was supported in part by the ``String Theory and Inflation'' Uncovering Excellence Grant of the University of Rome
``Tor Vergata", CUP E82L15000300005, and by the MIUR PRIN Contract
2015MP2CX4 ``Non-perturbative Aspects Of Gauge Theories And String''.
We thanks Michele Cicoli for the useful discussion.
P.C thanks ``Isola che non c'e' srl'' for the support.
\end{acknowledgments}




\nocite{*}

\bibliography{apssamp}

\begin{thebibliography}{10}
\bibitem{37} The topology of CY spaces is characterized by the two Hodge numbers $h^{1,1}$ and $h^{1,2}$.
\bibitem{41} Actually, one also has to include an uplifting term to get a Minkowski or a dS vacuum, see \cite{16}.
\end{thebibliography}

\end{document}